    \documentclass[aps,prb,twocolumn,groupedaddress,amsmath]{revtex4}
       \usepackage{bm}
    \usepackage{graphicx}
\newcommand\beq{\begin{equation}}
\newcommand\eeq{\end{equation}}
\newcommand\beqa{\begin{eqnarray}}
\newcommand\eeqa{\end{eqnarray}}
\newcommand{\dd}{\text{d}}
\newcommand{\ee}{\text{e}}
\newcommand{\nn}{\nonumber\\}
\newcommand{\rb}{\rho,\beta}

\newcommand{\pk}{\widetilde{\varphi}}
\newcommand{\hk}{\widetilde{h}}

\begin{document}
\title{Thermodynamic consistency between the energy and virial routes in the mean spherical approximation for soft potentials}
\author{Andr\'es Santos}
\email{andres@unex.es} \homepage{http://www.unex.es/fisteor/andres/}
\affiliation{Departamento de F\'{\i}sica, Universidad de
Extremadura, E-06071 Badajoz, Spain}
\date{\today}
\begin{abstract}
It is proven that, for any soft potential $\varphi(r)$ characterized
by a finite Fourier transform $\widetilde{\varphi}(k)$, the virial
and energy thermodynamic routes are equivalent if the Fourier
transform of the total correlation function divided by the density
$\rho$, $\widetilde{h}(k)/\rho$, is an arbitrary function of
$\rho\beta\widetilde{\varphi}(k)$, where $\beta$ is the inverse
temperature. This class of approximations includes the mean
spherical approximation as a particular  case.

\end{abstract}
\maketitle

\maketitle

Standard statistical-mechanical formulas relate the thermodynamic
properties of a fluid to the two-body interaction potential
$\varphi(r)$ and the associated radial distribution function
$g(r;\rho,\beta)$,\cite{BH76} where $\rho$ is the number density and
$\beta=1/k_BT$ is the inverse temperature.  When  an approximation
is used to get $g(r;\rho,\beta)$, there is in general no guarantee
that those  different formulas or routes are thermodynamically
self-consistent. In fact, some liquid state theories contain one or
more adjustable state-dependent parameters which are tuned to
achieve thermodynamic consistency between two or more routes. This
is the case, for instance, of the modified hypernetted-chain
closure, \cite{RA79} the Rogers--Young closure,\cite{RY84} the
Zerah--Hansen closure,\cite{ZH85} the self-consistent
Ornstein--Zernike approximation,\cite{HS77} the hierarchical
reference theory,\cite{PR95} Lee's theory based on the
zero-separation theorems, \cite{L95} the generalized mean spherical
approximation,\cite{W73} or the rational-function
approximation.\cite{YS91} It is then remarkable when an approximate
theory for $g(r;\rho,\beta)$ satisfies a condition of thermodynamic
consistency without being forced to do so. In this context, it is
perhaps not sufficiently well known that the hypernetted-chain (HNC)
integral equation provides thermodynamically consistent results
through the virial and energy routes, regardless of the potential
$\varphi(r)$.\cite{nBH76} More recently, Mladek \emph{et
al.}\cite{MKN06} have shown that the mean spherical approximation
(MSA) is exactly solvable for the Gaussian core model and have found
that the virial and energy routes to thermodynamics are equivalent
in that case as well.

The aim of this Note is to place Mladek \emph{et al.}'s
finding\cite{MKN06} in a broader context by proving the
thermodynamic consistency between the virial and energy routes for
(a) any ``soft'' potential and (b) within a class of approximations
that includes the MSA as a particular case.

The virial and energy routes to thermodynamics read\cite{BH76}
\beq
\frac{\beta p}{\rho}\equiv Z(\rho,\beta)=1-\frac{1}{2d}\rho\beta\int
\dd \mathbf{r}\, g(r;\rho,\beta)\mathbf{r}\cdot\nabla \varphi(r),
\label{1}
\eeq
\beq
u(\rho,\beta)= \frac{d}{2\beta}+\frac{1}{2}\rho \int \dd
\mathbf{r}\, g(r;\rho,\beta)\varphi(r),
\label{3}
\eeq
respectively, where $p$ is the pressure, $Z$ is the compressibility
factor, $d$ is the dimensionality of the system, and $u$ is the
internal energy per particle. The condition of thermodynamic
consistency between both routes is
\beq
\rho\frac{\partial}{\partial\rho}u(\rho,\beta)=\frac{\partial}{\partial
\beta}{Z(\rho,\beta)}.
\label{5}
\eeq

Let us now  consider an interaction potential verifying the boundary
conditions
\beq
\lim_{r\to 0} r^d \varphi(r)=0, \quad \lim_{r\to \infty} r^d
\varphi(r)=0.
\label{4}
\eeq
While the second condition means that the potential is sufficiently
short ranged, the first condition defines the kind of soft
potentials to be considered here. It includes bounded potentials
(such as the Gaussian core model\cite{SS97} or the penetrable sphere
model\cite{H57}), logarithmically diverging
potentials,\cite{LLWAJAR98} or even potentials diverging
algebraically as $\varphi(r)\sim r^{-n}$ with $n<d$. On the other
hand, conventional molecular models (such as hard spheres,
square-well fluids, and Lennard--Jones fluids) are excluded from the
class of potentials \eqref{4}. Equation \eqref{4} implies that
$\pk(0)=\text{finite}$, where the Fourier transform of the potential
is
\beq
\pk(k)=\mathcal{F}[\varphi(r)]=\int\dd\mathbf{r}\,
\ee^{-i\mathbf{k}\cdot\mathbf{r}}\varphi(r).
\label{6}
\eeq

Introducing the total correlation function $h(r;\rho,\beta)\equiv
g(r;\rb)-1$, Eqs.\ \eqref{1} and \eqref{3} can be rewritten as
\beq
Z(\rb)=1+\frac{1}{2}\rho\beta\pk(0)-\frac{1}{2d}\rho\beta I_v(\rb),
\label{7}
\eeq
\beq
u(\rb)=\frac{d}{2\beta}+\frac{1}{2}\rho\pk(0)+\frac{1}{2}\rho
I_e(\rb),
\label{8}
\eeq
where
\beqa
I_v(\rb)&\equiv &\int \dd \mathbf{r}\,
h(r;\rho,\beta)\mathbf{r}\cdot\nabla \varphi(r)\nn &=&
-\frac{1}{(2\pi)^{d}}\int\dd \mathbf{k}\,
\hk(k;\rho,\beta)\frac{\partial}{\partial\mathbf{k}}\cdot
\left[\mathbf{k} \pk(k)\right],
\label{9}
\eeqa
\beqa
I_e(\rb)&\equiv &\int \dd \mathbf{r}\, h(r;\rho,\beta) \varphi(r)\nn
&=& \frac{1}{(2\pi)^{d}}\int\dd \mathbf{k}\, \hk(k;\rho,\beta)
\pk(k).
\label{10}
\eeqa
In the last equalities of Eqs.\ \eqref{9} and \eqref{10},
$\hk(k;\rb)=\mathcal{F}[h(r;\rb)]$ and standard steps have been
followed. The consistency condition \eqref{5} becomes
\beq
\Delta(\rb)\equiv \frac{\partial}{\partial \rho}\left[\rho
I_e(\rb)\right]+\frac{1}{d} \frac{\partial}{\partial
\beta}\left[\beta I_v(\rb)\right]=0.
\label{11}
\eeq

Equation \eqref{11} is of course satisfied if the \emph{exact}
$\hk(k;\rb)$ is used to evaluate $I_v(\rb)$ and $I_e(\rb)$. On the
other hand, as proven below, the same happens with any
\emph{approximate} function $\hk(k;\rb)$ which, when divided by
$\rho$, depends on $\rho$, $\beta$, and $k$ through the scaled
variable $z\equiv \rho\beta\pk(k)$ only, i.e.,
\beq
\hk(k;\rb)={\rho}^{-1}F(z),\quad z\equiv \rho\beta\pk(k),
\label{12}
\eeq
where the function $F(z)$ does not need to be specified. The scaling
form \eqref{12} implies
\beq
\frac{\partial}{\partial \beta}\left[\beta
\hk(k;\rb)\right]=\hk(k;\rb)+\frac{\partial}{\partial
\rho}\left[\rho \hk(k;\rb)\right],
\label{13}
\eeq
\beq
\frac{\partial}{\partial \rho}\left[\rho
\hk(k;\rb)\right]\mathbf{k}\cdot\frac{\partial}{\partial\mathbf{k}}\pk(k)=\pk(k)\mathbf{k}\cdot\frac{\partial}{\partial\mathbf{k}}\hk(k;\rb).
\label{14}
\eeq
Applying Eq.\ \eqref{13} in the definition of $\Delta(\rb)$, Eq.\
\eqref{11}, one gets
\beqa
\Delta(\rb)&=&-\frac{1}{d(2\pi)^d}\int\dd \mathbf{k}\,\left\{
\hk(k;\rho,\beta)\frac{\partial}{\partial\mathbf{k}}\cdot
\left[\mathbf{k} \pk(k)\right]\right.\nn
&&\left.+\frac{\partial}{\partial \rho}\left[\rho
\hk(k;\rb)\right]\mathbf{k}\cdot\frac{\partial}{\partial\mathbf{k}}\pk(k)\right\}.
\label{15}
\eeqa
Finally,  use of Eq.\ \eqref{14} yields $\Delta(\rb)=0$, what proves
the thermodynamic consistency between Eqs.\ \eqref{1} and \eqref{3}
within the class of approximations \eqref{12}. Furthermore, $Z$ and
$u$ adopt the forms
\beq
Z(\rb)=1+\frac{1}{2}\pk(0)\alpha-\beta \zeta_v(\alpha),
\label{15b}
\eeq
\beq
\beta u(\rb)=\frac{d}{2}+\frac{1}{2}\pk(0)\alpha-\beta
\zeta_e(\alpha),
\label{16}
\eeq
where $\alpha\equiv\rho\beta$, $\zeta_v(\alpha)\equiv\rho
I_v(\rb)/2d$, and $\zeta_e(\alpha)\equiv-\rho I_e(\rb)/2$. Equation
\eqref{11} implies the relation
$\alpha\zeta_e'(\alpha)=\dd\left[\alpha\zeta_v(\alpha)\right]/\dd\alpha$,
where $\zeta_e'(\alpha)\equiv\dd\zeta_e(\alpha)/\dd\alpha$.

It must be noted that the compressibility route is inconsistent with
the virial and energy routes, i.e., $[1+\rho\hk(0;\rb)]^{-1}\neq
\partial(\rho Z)/\partial \rho$, for approximations of the form \eqref{12}.
While the (reduced) isothermal compressibility
$1+\rho\hk(0;\rb)=1+F\left(\pk(0)\alpha\right)$ depends on $\rho$
and $\beta$ through the product $\alpha=\rho\beta$ only,  Eq.\
\eqref{15b} yields  $\partial(\rho Z)/\partial
\rho=1+\pk(0)\alpha-\beta\alpha\zeta_e'(\alpha)$, the last term
depending on both $\alpha$ and $\beta$.

The MSA for soft potentials\cite{MKN06}  consists of assuming the
random-phase approximation $c(r;\rb)=-\beta \varphi(r)$ for the
direct correlation function at any distance. The Ornstein--Zernike
relation\cite{BH76} then gives Eq.\ \eqref{12} with $F(z)=-z/(1+z)$,
so that this version of the MSA is thermodynamically consistent
through the virial and energy routes. However, they are inconsistent
with the the compressibility route: $[1+\rho\hk(0;\rb)]^{-1}-
\partial(\rho Z)/\partial \rho=\beta\alpha\zeta_e'(\alpha)$.

The work presented here and elsewhere\cite{S05} might have a
didactic value in showing that relatively simple mathematics allows
one to check the  thermodynamic self-consistency between the virial
and energy routes in some important special cases.

 This research
has been supported by the Ministerio de Educaci\'on y Ciencia
(Spain) through Grant No. FIS2004-01399 (partially financed by FEDER
funds).

\end{document}